\documentclass[floats,preprint,aps,amssymb,prb,secnumarabic,eqsecnum]{revtex4}

\begin{document}


Comment on ``Failure of the Jarzynski identity for a simple quantum system.''
\vspace{0.5cm}
\begin{center}
  Shaul Mukamel \\ Department of Chemistry, University of California, Irvine \\
  Irvine, CA 92697-2025 \\ smukamel@uci.edu 
\end{center}
\vspace{0.5cm}
The article of Engel and Nolte [1] is based on an incorrect definition of work. 
In fact, the Jarzynski identity does hold. 
The energy and work distribution of a driven, but otherwise isolated, quantum system may be calculated using a time-dependent 
(adiabatic) basis which diagonalizes the Hamiltonian at a given time. 
The general proof of the Jarzynski relation for an arbitrary driving protocol was given in ref. [2]. 
Let us consider the special case discussed in [1] where we suddenly switch the Hamiltonian $H_0$ into $H_1$. 
We denote the corresponding partition functions $Z_0$ and $Z_1$. Consider a quantum trajectory 
where the system starts in an eigenstate $\varphi_j$ of $H_0$ with eigenvalue $\varepsilon_j$, and ends up in an eigenstate 
$\Psi_n$ of $H_1$ with eigenvalue $\varepsilon_n$. 
The probability of this trajectory is 
\begin{equation}
P_{nj} = \frac{\exp(-\beta \varepsilon_j)}{Z_0} | c_{nj} |^2,
\end{equation}
where $c_{nj}$ are the expansion coefficients of $\varphi_j$ in the new basis $\varphi_j$ = $\sum_n c_{nj} \Psi_n$, with 
$\sum_j | c_{nj} |^2$ = 1.
Since there is no bath, the work done on the system is $W_{nj}$ = $\varepsilon_n$ - 
$\varepsilon_j$.  Using these definitions we immediately recover the Jarzynski relation 
\begin{equation}
\langle e^{-\beta W} \rangle \equiv \sum_{jn} e^{-\beta W_{nj}} P_{nj} = \frac{Z_1}{Z_0} \equiv \exp(-\beta \Delta F) 
\end{equation} 

\vspace{1cm}
Acknowledgments
\vspace{0.5cm}

The support of the Chemical Sciences, Geosciences and Biosciences Division, Office of Basic Energy Sciences, 
Office of Science, and U.S. Department of Energy is gratefully acknowledged.

\vspace{1cm}
[1] Engel, A., Nolte, R. cond-mat/0612527v1 (2006)

[2] Mukamel, S. Phys. Rev. Lett. 90, 170604 (2003)

\end{document}